\documentclass[prl,aps,twocolumn,epsfig]{revtex4}
\usepackage{epsfig}
\usepackage{amsmath}
\usepackage{amssymb}
\usepackage{bm}
\usepackage{dcolumn}

\newcommand{\beq}{\begin{equation}}
\newcommand{\eeq}{\end{equation}}
\newcommand{\bea}{\begin{eqnarray}}
\newcommand{\eea}{\end{eqnarray}}
\newcommand{\bei}{\begin{itemize}}
\newcommand{\eei}{\end{itemize}}

\newcommand{\br}{{\mathbf{r}}}
\newcommand{\bk}{{\mathbf{k}}}
\def\half{\frac{1}{2}}

\def\ket#1{\left|{#1}\right>}

\begin{document}

\title{Paired composite fermion phase of quantum Hall bilayers at $\nu=\frac{1}{2} + \frac{1}{2}$}

\author{Gunnar M\"{o}ller$^1$,  Steven H. Simon$^2$, and Edward H. Rezayi${}^{3}$}
\affiliation{$^1$Theory of Condensed Matter Group, Cavendish Laboratory, J.~J.~Thomson Ave., Cambridge CB3~0HE, UK\\
${}^2$Bell Laboratories, Alcatel-Lucent, 600 Mountain Road, Murray Hill, New Jersey 07974
\\
${}^3$Department of Physics, California State University, Los Angeles, California 90032 }

\begin{abstract}
We provide numerical evidence for composite fermion pairing in quantum Hall bilayer systems at filling $\nu=\half+\half$ for intermediate spacing between the layers. We identify the phase as $p_x  + i p_y$ pairing, and construct high accuracy trial wavefunctions to describe the groundstate on the sphere.  For large distances between the layers, and for finite systems, a competing ``Hund's rule" state, or composite fermion liquid, prevails for certain system sizes.  
\end{abstract}
\date{April 8th, 2008}
\maketitle

The bilayer quantum Hall system at filling fraction $\nu=\half +
\half$ has been an active topic of research for over a decade \cite{DasSarmaPerspectives,DLReviewNature}.  In the limit of small layer spacing $d$ there is a superfluid phase, whereas in the limit of infinite $d$ the two layers form
independent compressible composite fermion (CF) seas \cite{Heinonen}.   Much less well understood, however, is the nature of the state at intermediate $d$.  Even in simplified models with spinless electrons at zero temperature, no disorder, and no tunneling between layers (simplifications we will adopt throughout this paper), there are a wide range of theoretical calculations, predictions, and scenerios \cite{OtherTheory,KimNayak01,Bonesteel,PRLSimon03} attempting to address the transition.  One of the most interesting possibilities \cite{Bonesteel} is that, even for very weak interaction between layers, the interaction between the two CF seas may cause BCS pairing between the layers to form a superconducting state: an ``interlayer paired CF-BCS state".  Unfortunately, this suggestion was based on a highly approximate perturbative Chern-Simons approach whose validity at higher order is impossible to test.   In the current paper we provide highly accurate numerical evidence that for a range of distances between the layers, the groundstate is indeed such a CF-BCS state.   We further show that the pairing symmetry is in the $p_x + i p_y$ channel in contrast to the theoretical arguments of Ref.~\cite{KimNayak01} which suggest $p_x - i p_y$.  We construct and numerically verify the first explicit wavefunctions describing this phase.  For finite systems for very large $d$ we also identify the groundstate to be a ``Hund's rule" or composite fermion liquid state.  For certain shell fillings (number of particles) this is precisely the weak-interaction limit of the CF-BCS state, whereas for other fillings, it is a distinct state.  Finally, we argue that, when the two possibilities are distinct, the range in $d$ for which the Hund's rule state prevails over the CF-BCS state becomes smaller as we go to larger systems while 
the paired state becomes more predominant.


Since many of our technical details are similar to that of Ref.~\cite{MollerSimon}, where single-layer pairing is considered, we will be brief with our discussion.   We start by considering the wavefunction $\prod_\bk \left( 1 + g_\bk \,e^{i\varphi} \, c^\dagger_{\bk \uparrow}
  c^\dagger_{-\bk \downarrow}\right)\ket{0}$ which represents a
  BCS paired wavefunction \cite{deGennesSupra} for a bilayer spinless fermion system in zero magnetic field (where $\uparrow, \downarrow$ represent the layer index).
Fourier transforming with respect to $\varphi$ we project to precisely $N/2$ fermions ($N$ assumed even) in each layer \cite{deGennesSupra}, yielding
$\Psi_\text{BCS} = \det \left[ G(\br_{i,\uparrow}-\br_{j,\downarrow}) \right]$ where
$G(\br_{i,\uparrow}-\br_{j,\downarrow})$ is an $N/2$ by $N/2$ matrix with indices $i$ and $j$.  $G(\br_{i,\uparrow}-\br_{j,\downarrow})$, the wavefunction of a pair, can be written in terms of $g_\bk$ as
\begin{equation} \label{eq:gFourier} G(\br_{i,\uparrow}-\br_{j,\downarrow}) = \mbox{$\sum_\bk$} \, g_\bk \phi_\bk(\br_{i,\uparrow}) \,
\phi_{-\bk}(\br_{j,\downarrow})
\end{equation}
where $\phi_{\bk}(\br) = e^{i \bk \cdot \br}$ are the simple single fermion plane wave orbitals, such that Eq.~\ref{eq:gFourier} is just a Fourier transform.  BCS theory is fundamentally variational \cite{deGennesSupra}, and one uses a form of $g_\bk$ which minimizes the energy given a particular inter-particle interaction.  The ``symmetry" or relative angular momentum of the pairing wavefunction is determined by the phase winding \cite{endnote1} of $G$.  If the phase of $G(\br_{i,\uparrow}-\br_{j,\downarrow})$ wraps by $2 M \pi$ as $\br_{i,\uparrow}$ is taken clockwise around $\br_{j,\downarrow}$, we say the wavefunction has angular momentum $M$ (or $M$-wave symmetry) where $M=0,\pm 1$ are also known as $s,p_x \pm i p_y$ respectively.    (While some prior literature uses ``$p_x + i p_y$" imprecisely to denote either chirality).  Note that the limit of noninteracting fermions can be achieved with this form by taking $g_\bk > 0$ for all $|\bk| < k_F$ and  $g_\bk = 0$ otherwise (with $k_F$ the Fermi momentum).  Thus, the BCS paired state can be deformed smoothly into a noninteracting Fermi liquid (a limit point outside of the superconducting phase).

Our calculations are performed on a spherical geometry.    We remind the reader that with a magnetic monopole of flux $2 q$ at the center of the sphere (with $q$ half integer), the single particle eigenstates \cite{WuYang} are the spherical monopole harmonics $Y^q_{l,m}(\br)$ where $l=q,q+1, \ldots$, and $m = -l, -l+1, \ldots, l$ and the energy of these states depend on $l$ only.     From these orbitals we construct a general pair wavefunction
\begin{equation}
\label{eq:Gsphere} G(\br_{i,\uparrow},\br_{j,\downarrow}) = \mbox{$\sum_{l,m}$} g_l (-1)^{q + m} Y^q_{l,m}(\br_{i,\uparrow}) \,
Y^q_{l,-m}(\br_{j,\downarrow})
\end{equation}
where again the $g_l$ are variational parameters that control the (radial) shape of the pair wavefunction.   It is easy to show that this pairing wavefunction has $M$-wave symmetry with $M=2q$.   Indeed, one can show that in order to have a BCS groundstate with $M$-wave symmetry on the sphere with no  vortex defects, one must have a monopole of flux $M$ at the center of the sphere,  and the pairing form is always expressible in the form of Eq.~\ref{eq:Gsphere}.

To convert our simple BCS wavefunction into a paired composite fermion \cite{Heinonen,JainKamilla97} wavefunction appropriate for modeling the bilayer quantum Hall state at $\nu=\half+\half$, we must multiply by Jastrow factors such that electrons ``see'' flux attached to electrons only within the same layer.  To achieve this, we replace the single particle orbitals $Y^q_{l,m}$ with composite fermion orbitals defined as $\tilde Y^q_{l,m}(u_{i,\alpha},v_{i, \alpha}) = {\cal P} [ J_{i,\alpha} Y^q_{l,m}(\br_{i,\alpha})] $ where $u_{i,\alpha},v_{i,\alpha}$ is the usual spinor representation of the coordinate $\br_{i,\alpha}$ on the sphere, $\alpha = \uparrow, \downarrow$ is the layer index, $J_{i,\alpha} = \prod_{j \neq i} (u_{i,\alpha} v_{j, \alpha} - v_{i,\alpha} u_{j, \alpha})$ is the composite fermionization factor, and $\cal P$ is the projection to the lowest Landau level.  Note that in our present notation, the Jastrow factor $J$ is absorbed within the composite fermion wavefunction $\tilde Y$ 
(as opposed to Refs.~\cite{JainKamilla97}).     As usual, $\tilde Y^q_{l,m}(u_{i,\alpha},v_{i, \alpha})$ is implicitly a function of all the particle coordinates, although we only denote it explicitly as a function of particle $i$.

We substitute these composite fermion orbitals $\tilde Y$ in place of the orbitals $Y$ in Eq.~\ref{eq:Gsphere} to generate the pairing wavefunction which we will correspondingly call $\tilde G$.   Our trial wavefunction for the bilayer quantum Hall state is then $\Psi = \det \left[ \tilde G(\br_{i,\uparrow},\br_{j,\downarrow}) \right]$, completely analogous to the case of simple BCS theory discussed above.  The generated trial wavefunction (for $N/2$ electrons in each layer) occurs on the sphere with a monopole flux $N_{\phi} = 2(N/2-1) + M$  for the case of $M$-wave symmetry (this ``shift" of $M$ is caused by the small addition of $M$ flux quanta necessary to avoid having vortices, as discussed above).  The different possible pairing symmetries can then be easily identified by their shifts.

\begin{table}
\begin{tabular}{cccccccccccc}
$M$ & $N$ & ~~~ $d =$ \!\!\! &1.0&&1.5&&2.0&&2.5&&3.0\\
\hline
-1 &10&& \textbf{1}&&\textbf{2}&&\textbf{2}&&\textbf{2}&&\textbf{1}\\
 -1 &12 && 0.09687&&0.1519&&0.1739&&0.1813&&0.1844\\
 -1 &14 && \textbf{2} &&\textbf{1}&&\textbf{1}&&\textbf{1}&&0.0004\\
-1 &16&& 0.0074 &&\textbf{3}&&0.0008&&0.0019&&0.0017\\
\hline
0&10 && 0.0222&&0.0133&&0.0108&&0.0077&&0.0054\\
0 &12 && \textbf{1}&&\textbf{1}&&0.0019&&0.0040&&0.0036\\[0pt]
 0 &14&& \textbf{2} &&\textbf{2}&&\textbf{2}&&\textbf{1}&&0.0004\\
0 &16&& \textbf{3} &&\textbf{1}&&\textbf{1}&&\textbf{1}&&\textbf{1}\\
\hline
+1&10 && 0.1517&&0.1070&&0.0508&&0.0254&&0.0138\\
+1 &12 && 0.1438&&0.1048&&0.1466&&0.1486&&0.1495\\[0pt]
 +1 &14 && 0.1373&&0.0828&&0.0334&&0.0151&&0.0080\\
+1&16 && 0.1316&&0.0754&&0.0246&&0.0089&&0.0042\\
\hline\hline
\end{tabular}
\caption{Data for bilayer electrons on a sphere near $\nu=\half+\half$ for various interlayer spacings $d$ (in units of the magnetic length $\ell_0$).  The table shows energy gaps (in units of $e^2/\ell_0$) for cases where the groundstate is angular momentum zero (non-bold), and angular momentum $L$ of the groundstate (bold) when it is not zero.   Data is shown with $N/2$ electrons per layer with flux $N_{\phi} = N - 2 + M$ for $M=-1,0,+1$ corresponding to the pairing symmetries $p_x - i p_y$, $s$, and $p_x + ip_y$.  Only the case of $p_x + i p_y$ consistently shows a strong gap at zero angular momentum characteristic of a quantum Hall state.   Data for $M=\pm 2$ (not shown) also does not suggest a quantum Hall state.   We thus identify $p_x + i p_y$ as the most likely pairing symmetry. Note that the strong gap for $N=12$ in the $M=-1,+1$ case both correspond to filled shells of composite fermions. In non-filled shells, the gap drops with increasing $d$ as the interlayer interaction is reduced.}
\label{tab}
\end{table}

Our numerical analysis is founded on 
exact diagonalizations on the sphere.  The inter-electron interaction is taken to be $V(\br_{i,\alpha}, \br_{j,\beta}) = e^2/r$ for $\alpha = \beta$ and $e^2/\sqrt{r^2 + d^2}$  for $\alpha \neq \beta$ where $d$ represents the ``spacing" between layers ($\alpha, \beta = \uparrow, \downarrow$ are the layer indices). We have defined the magnetic length to be unity, and $r$ is the chord-distance between $\br_i$ and $\br_j$.   In table \ref{tab} we examine the stability of the bilayer system at different shifts (different values of the magnetic monopole flux).  We recall that quantum Hall states correspond to groundstates at angular momentum $L=0$ which have a strong gap.  While it is obvious that there should be a gap and an $L=0$ groundstate when each individual layer is gapped with an $L=0$ groundstate (which is the case for $N=12$ for both $M=+1,-1$), it is less trivial in the other cases where the individual layers have $L \neq 0$ groundstates.   We find that the $M=+1$ case has an $L=0$ groundstate with a strong gap for all values of the interlayer spacing $d$, 
while this is not the case for other values of $M$ ($M=0,\pm 1, \pm 2$ have been examined).  We thus suspect that if a BCS paired state of CF's exists, it is in the $p_x + i p_y$ channel ($M=+1$).    We will thus focus on trial CF-BCS wavefunctions with this pairing symmetry.

\begin{figure}
\begin{center}
\includegraphics[width=0.78\columnwidth]{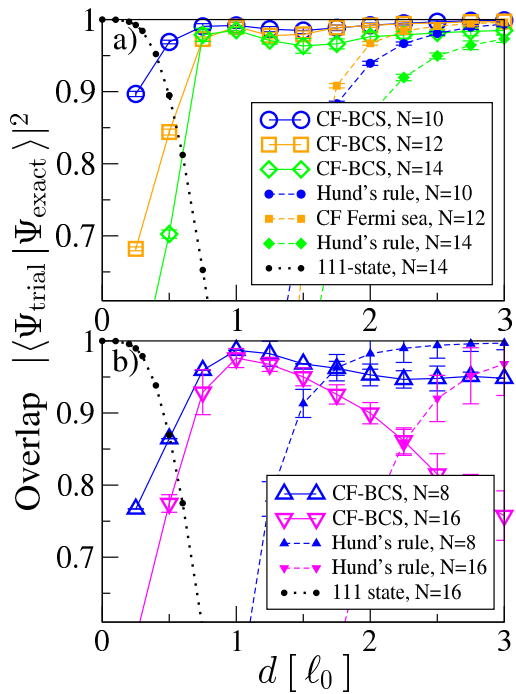}
\end{center}
\caption{(color online) Overlaps of our trial CF-BCS states with the exact groundstate on the bilayer sphere (solid), overlaps of the Hund's rule or CF-Fermi sea state with the exact groundstate (dashed) and overlaps with the 111-state (dotted).  Error bars are Monte-Carlo error.  In (a) data is shown for $N=10,12,14$ where the large $d$ limit of the Hund's rule and CF-BCS state become identical.   Note the very high accuracy of these trial wavefunctions for all $d \gtrsim 1$.   In (b) for $N=8,16$ the two limits are inequivalent.  In this case it is clear that the Hund's rule state becomes more accurate at large $d$, but at smaller $d$ there is a regime where the CF-BCS state prevails. The $111$-state is accurate only at $d\lesssim 0.5\ell_0$.
} \label{fig:overlaps1}
\end{figure}

Using the above described approach, we generate trial CF-BCS wavefunctions with $p_x + i p_y$ pairing.  Calculations are performed by Monte-Carlo, and at each value of the interlayer spacing $d$, the shape of the pairing wavefunction is optimized by varying the parameters $g_l$ to maximize the overlap with the exact groundstate (See Ref.~\cite{MollerSimon} for details of the calculational scheme).   Very good agreement with the exact diagonalization is obtained by only varying the first few such parameters.   In Fig.~\ref{fig:overlaps1}a we show overlaps of our trial wavefunctions with the exact groundstate for several different sized systems ($N=10,12,14$) over a range of $d$.   For $d \gtrsim 1$ the overlaps of the trial state with the exact groundstate are excellent.  Since these wavefunctions are variational, one might worry that a large number of variational parameters is required to obtain such good agreement.  However, the dimension $D$ of the $L=0$ Hilbert space is $D=38,252,1559$ for $N=10,12,14$, and we have used only $3,4$ and $4$ variational parameters $g_l$ respectively, so these agreements are highly significant.

Also shown in Fig.~\ref{fig:overlaps1}a is the overlap of the exact groundstate with the CF-Fermi sea trial wavefunction, as well as that with the $111$-state.   For the $N=12$  case, the CF-Fermi sea state is uniquely defined as filling two CF shells in each layer  (at this flux, the $p^{th}$ CF shell has $2p$ orbitals)\cite{Heinonen,JainKamilla97}.  At large $d$ this CF-Fermi sea has virtually perfect overlap with the exact groundstate. Furthermore, the bilayer CF-Fermi sea in this case is simply a limit of the CF-BCS wavefunction (as described above).
However, at smaller $d$, where the CF-BCS wavefunction achieves almost perfect agreement with the exact groundstate, the CF-Fermi sea has very poor overlap.

For values of $N$ which are not filled shell situations, the CF-Fermi sea needs to be more carefully defined.  We recall that for a single-layer system (at least for small systems), the groundstate of a partially filled shell satisfies Hund's rule by filling degenerate CF-orbitals to maximize the total angular momentum \cite{ReadRezayiCF}.
We thus propose that the groundstate of the bilayer at very large $d$ is obtained by combining two single-layer states, each of which satisfies Hund's rule individually, into an overall angular momentum singlet  (adding up the wavefunctions of the two individual layers into the unique $L_\text{total}=0$ state using appropriate Clebsch-Gordan coefficients).  We call this construction the ``Hund's rule" bilayer state.   We see in Figs.~\ref{fig:overlaps1}a and \ref{fig:overlaps1}b that the Hund's rule state always becomes extremely accurate at large $d$.  In cases where there is either one CF in the valence shell in each layer (such as $N=14$) or one CF-hole in the valence shell in each layer(such as $N=10$), it can be shown that the Hund's rule state is again precisely the large $d$ limit of the CF-BCS state.  Indeed, the Hund's rule and CF-BCS state become identical in this limit (and both become almost perfect).   However, in other cases where there is an incompletely filled shell ($N=8,16$ shown in Fig.~\ref{fig:overlaps1}b), the Hund's rule state cannot be written as a limit of the CF-BCS state.  To understand that these must be different, we note that for the Hund's rule state, each layer is an eigenstate of $L^2$, whereas for the BCS state this is only true in the limit where the pairing becomes infinitely weak {\it and} when it is a filled shell situation, or there is a single electron or single hole in each valence shell.

As seen in Fig.~\ref{fig:overlaps1}b, for $N=8,16$, the Hund's rule state becomes extremely accurate in the large $d$ limit, and the CF-BCS state becomes inaccurate. Remarkably, as we go to smaller $d$, the CF-BCS state again becomes extremely accurate whereas the Hund's rule state fails.  This is an extremely important result: even when the large $d$ limit is not of the CF-BCS pairing form, when the interaction between layers is increased, the pairing form again becomes accurate.   (For $N=8,16$ 
the dimension of the $L=0$ Hilbert space is $D=12,12774$ 
and we have used $2,4$ variational parameters, so again the agreement is significant. 
In all cases, the $111$-state which describes the interlayer coherent phase is accurate only at small $d\lesssim 0.5\ell_0$, where its overlap with the exact groundstate plummets and the CF-BCS states become the best trial states. Thus, we argue that there is a region of intermediate $d$ where a CF-BCS phase is the groundstate.

\begin{figure}
\begin{center}
\includegraphics[width=0.78\columnwidth]{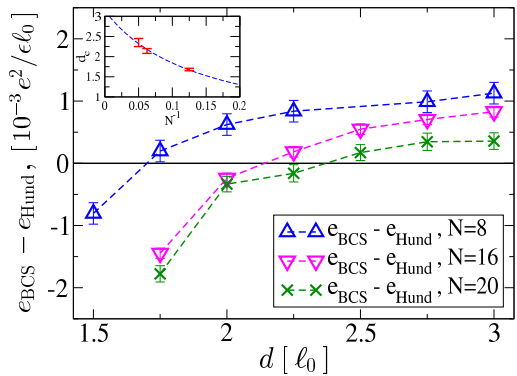}
\end{center}
\caption{(color online) Energy difference between CF-BCS wavefunction and Hund's rule wavefunction as a function of layer spacing $d$ for shell fillings where CF-BCS and Hund's rule are distinct states in the large $d$ limit. We show $N=8,16,20$.   Note that as $N$ grows, the range of $d$ where the CF-BCS wavefunction is a better trial state extends to larger $d$. The inset shows $d_c$, the value of $d$ where $e_\text{BCS}=e_\text{Hund}$ to illustrates how it scales with $N^{-1}$; lines are guides to the eye.
}
\label{fig:hund}
\end{figure}

We further conjecture that as we go to larger systems, the shell-filling effects, and Hund's rule, should become less important whereas the pairing effects will remain the same strength.   This conjecture is simply based on the fact that only $\sim \sqrt{N}$ particles are in the valence shell, whereas all particles within some gap energy of the Fermi surface (a number $\sim N$) contribute to pairing.  To make this statement more concrete we examine the range for which the CF-BCS wavefunction provides a better trial state than the Hund's rule state.  In Fig.~\ref{fig:hund} we show the energy difference per particle between the two trial states as a function of system size and layer separation. To differentiate between the two possibilities most clearly, we have only shown cases ($N=8,16,20$) where the large $d$ limit of the CF-BCS state is distinct from the Hund's rule state. The variational CF-BCS wavefunction for $N=20$ electrons was obtained using an energy minimization technique \cite{LongBilayer}.  The figure shows that with increasing $N$, the CF-BCS state becomes more accurate out to somewhat larger $d$. While our data strongly suggests that pairing survives in the thermodynamic limit it cannot establish that it extends to arbitrary weak interactions, as suggested in Ref.~\cite{Bonesteel}, in this limit (see inset).

For large $d$, it is difficult to be confident of extrapolation to the thermodynamic limit (small energy differences).  However, for intermediate $\ell_0 \lesssim d \lesssim 2 \ell_0$, our data (table \ref{tab}) clearly shows a strong gap which persists to large $N$.  At these values of $d$, it is also clear from our numerics that the CF-BCS state provides an accurate trial wavefunction which is not in the 111 phase.  We can thus
conclude that a CF-BCS phase, contiguous to the 111 phase, does exist for a range of intermediate $d$.


We emphasize that the CF-BCS trial wavefunctions contain interlayer correlations via the interlayer pairing function $G$, whereas the intra-layer correlations result from the CF Jastrow factors.  As such, these trial wavefunctions should not have the same interlayer coherence as the 111 wavefunction.
Still, our CF-BCS states should have quantized Hall drag exactly like the 111 phase, at least at zero temperature (see  \cite{KimNayak01,ReadGreen}), although they should not display resonant interlayer tunneling.  Finally, we note that, unlike the 111 phase, the CF-BCS state will be strongly destabilized by layer imbalance (analogous to spin imbalance for usual BCS states).  

In a forthcoming paper \cite{LongBilayer} we will 
discuss the overall phase diagram including the regime of very small $d$.   In brief,
we find a second order phase transition near $d \gtrsim 1$ from the CF-BCS phase at larger $d$, which has zero 111 order parameter (OP), to a phase with nonzero 111 OP at smaller $d$. Note that the BCS OP and the 111 OP are distinct.   We find that both phases as well as the transition can be very well described in the language of \cite{PRLSimon03} where CF's mix with composite bosons (CB's) and the presence of CB's yields nonzero 111 OP.   Our work further suggests that there may be a region of intermediate $d$ where both OPs (BCS and 111) coexist, although the results of \cite{LongBilayer} are not definitive in this respect.   
Should this coexistence occur, we argue \cite{LongBilayer} that $p_x+ip_y$ is the only pairing channel compatible with such coexistence.


To summarize, we have shown compelling numerical evidence of $p_x + i p_y$ pairing of composite fermions at intermediate layer spacings $d$ for quantum Hall bilayers at $\nu=\half+\half$.    We have proposed specific forms for the actual wavefunctions that show excellent overlap with the results of exact diagonalizations.   While CF pairing had been theoretically proposed earlier \cite{Bonesteel}, other phases had also been advocated \cite{KimNayak01,OtherTheory} and there has previously been no compelling numerical evidence to distinguish the possibilities.  Further we show that for finite size systems at very large $d$, a Hund's rule state is the groundstate.  However, as we go to larger system sizes, the CF-BCS (paired) state extends to larger $d$.

\acknowledgments

The authors acknowledge helpful discussions with N.~R.~Cooper and J.~P.~Eisenstein,
and thank the Aspen Center of Physics for hospitality.
G.M.~acknowledges support by EPSRC Grant No.~GR/S61263/01.
E.~R. is supported by DOE grant DE-FG03-02ER-45981.


\begin{thebibliography}{50}

\bibitem[{\mbox{Das Sarma} and Pinczuk(1997)}]{DasSarmaPerspectives}
S.~\mbox{Das Sarma} and A.~Pinczuk (eds.), \emph{Perspectives in Quantum Hall
  effects} (Wiley, New York, 1997).

\bibitem[{Eisentein and MacDonald(2004)}]{DLReviewNature}
See for example, J.~P. Eisentein and A.~H. MacDonald, Nature \textbf{432}, 691 (2004), and references therein.

\bibitem[{Heinonen(1998)}]{Heinonen} See, for example, O.~Heinonen (ed.), \emph{Composite Fermions} (World Scientific, 1998), for a
  review of CF's.


\bibitem[{Simon \emph{et~al.}(2003)Simon, Rezayi, and Milovanovic}]{PRLSimon03}
S.~H. Simon, E.~H. Rezayi, and M.~V. Milovanovic, Phys.\ Rev.\ Lett.
  \textbf{91}, 046803 (2003).

\bibitem[{Kim, Nayak, Demler, Read, and Sarma}]{KimNayak01}
Y.~B. Kim, C.~Nayak, E.~Demler, N.~Read, and S.~DasSarma, Phys.\ Rev.\ B
  \textbf{63}, 205315 (2001).




\bibitem[{Bonesteel(1993)}]{Bonesteel}
N.~E. Bonesteel, Phys.\ Rev.\ B \textbf{48}, 11484 (1993); N.~E. Bonesteel, I.~A. McDonald, and C.~Nayak, Phys.\ Rev.\ Lett. \textbf{77},
  3009 (1996).


\bibitem{OtherTheory}
For physics of intermediate $d$, see e.g., J.~Schliemann, S.~M. Girvin, and A.~H. MacDonald, Phys.\ Rev.\ Lett. \textbf{86}, 1849 (2001);  Y.~N. Joglekar and A.~H. MacDonald, Phys. Rev. B  \textbf{64}, 155315 (2001); {\it ibid} \textbf{65}, 235319 (2002); K.~Nomura and D.~Yoshioka, Phys.\ Rev.\ B \textbf{66}, 153310 (2002); J.~Schliemann, Phys.\ Rev.\ B \textbf{67}, 035328 (2003); K.~Park, Phys.\ Rev.\ B \textbf{69}, 045319 (2004);
N.~Shibata and D.~Yoshioka, J. Phys. Soc. Jpn. \textbf{75}, 043712 (2006);  J.~Ye, Phys.~Rev.~Lett. \textbf{97}, 236803 (2006); J. Ye and L.~Jiang, Phys.~Rev.~Lett. \textbf{98}, 236802 (2007) M.~V. Milovanovic, Phys.~Rev.~B \textbf{75}, 035314 (2007);



\bibitem{MollerSimon} G.~M\"{o}ller and S.~H.~Simon,  Phys.\ Rev.\ B. \textbf{77}, 075319 (2008).



\bibitem[{de~Gennes(1966)}]{deGennesSupra}
P.~G. de~Gennes, \emph{Superconductivity of Metals and Alloys} (Benjamin, New
  York, 1966).


\bibitem{endnote1} We focus on cases where there are no gapless excitations, which excludes symmetries such as $p_x$.

\bibitem{WuYang} T.~T.~Wu and C.~N.~Yang, Nucl.~Phys.~\textbf{B107}, 365 (1976).

\bibitem{JainKamilla97} J.~K. Jain and R.~K. Kamilla, Phys.\ Rev.\ B \textbf{55}, R4895 (1997); see also
 G.~M\"oller and S.~H. Simon, Phys.~Rev.~B \textbf{72}, 045344 (2005).



\bibitem{ReadRezayiCF}  E. Rezayi and N. Read, Phys.~Rev.~Lett. \textbf{72}, 900 (1994).

\bibitem{LongBilayer}
G.~M\"oller, S.~H. Simon, E.~H. Rezayi, to be published.

\bibitem{ReadGreen}
N.~Read and D.~Green, Phys.\ Rev.\ B \textbf{61}, 10267 (2000).



\end{thebibliography}
\end{document}